# TrajVis: a visual clinical decision support system to translate artificial intelligence trajectory models in the precision management of chronic kidney disease

Zuotian Li, MS[1,2], Xiang Liu, PhD [1], Ziyang Tang, PhD[3], Pengyue Zhang, PhD[1], Nanxin Jin, MS[1,3], Michael T. Eadon, MD[4], Qianqian Song, PhD[5*], Yingjie Chen, PhD[2,*], Jing Su, PhD[1,6, *]

[1]Department of Biostatistics and Health Data Science, Indiana University School of Medicine, Indiana, USA;

[2]Department of Computer Graphics Technology, Purdue University, Indiana, USA;

[3]Department of Computer and Information Technology, Purdue University, Indiana, USA;

[4]Department of Medicine, Indiana University School of Medicine, Indiana, USA;

[5]Department of Health Outcomes and Biomedical Informatics, College of Medicine, University of Florida, Florida, USA;

[6]Gerontology and Geriatric Medicine, Wake Forest School of Medicine, North Carolina, USA

**Correspondences:**

Jing Su

Department of Biostatistics & Health Data

Indiana University School of Medicine

Indianapolis, IN 46202

317-278-6451

su1@iu.edu

Yingjie Chen

Department of Computer Graphics Technology

Purdue University

West Lafayette, IN, 47907

765-494-1454

victorchen@purdue.edu

Qianqian Song

Department of Health Outcomes and Biomedical Informatics

College of Medicine

University of Florida

Gainesville, FL, 32611

qsong1@ufl.edu

**Abstract**

**Objective:** Our objective is to develop and validate TrajVis, an interactive tool that assists clinicians in using artificial intelligence (AI) models to leverage patients' longitudinal electronic medical records (EMR) for personalized precision management of chronic disease progression.

**Methods:** We first perform requirement analysis with clinicians and data scientists to determine the visual analytics tasks of the TrajVis system as well as its design and functionalities. A graph AI model for chronic kidney disease (CKD) trajectory inference named DEPOT is used for system development and demonstration. TrajVis is implemented as a full-stack web application with synthetic EMR data derived from the Atrium Health Wake Forest Baptist Translational Data Warehouse and the Indiana Network for Patient Care research database. A case study with a nephrologist and a user experience survey of clinicians and data scientists are conducted to evaluate the TrajVis system.

**Results:** The TrajVis clinical information system is composed of four panels: the Patient View for demographic and clinical information, the Trajectory View to visualize the DEPOT-derived CKD trajectories in latent space, the Clinical Indicator View to elucidate longitudinal patterns of clinical features and interpret DEPOT predictions, and the Analysis View to demonstrate personal CKD progression trajectories. System evaluations suggest that TrajVis supports clinicians in summarizing clinical data, identifying individualized risk predictors, and visualizing patient disease progression trajectories, overcoming the barriers of AI implementation in healthcare.

**Conclusion:** TrajVis bridges the gap between the fast-growing AI/ML modeling and the clinical use of such models for personalized and precision management of chronic diseases.

## 1. Introduction

Clinicians rely on clinical indicators to monitor the progression of chronic diseases. For example, according to the KDIGO (Kidney Disease: Improving Global Outcomes)[1, 2] and the NKF-KDOQI (National Kidney Foundation Kidney Disease Outcomes Quality Initiative) guidelines[3, 4], clinicians use estimated glomerular filtration rate (eGFR) and urinary albumin/creatinine ratio to monitor the progression of chronic kidney disease (CKD). However, chronic diseases are complex and heterogeneous. Relying on a few clinical indicators to monitor complex diseases raises two issues: delayed identification of progression, which leads to missing the opportunity window for intervention, and overdiagnosis due to the fluctuation of the key clinical indicators[4].

In recent years, progress has been made in implementing artificial intelligence (AI) algorithms into hospital care to predict and detect clinical status changes like acute kidney injury[5, 6]. These AI algorithms leverage the frequent measurements of clinical indicators in the inpatient electronic medical records (EMR) to predict acute disease risk during hospitalization. However, the AI modeling of progressive diseases like CKD depends on longitudinal EMR data. Such data is known for greater data missingness, irregular follow-up, and informative censoring, which challenges the clinical use of AI-based prediction models[7].

AI models recognize latent patterns in complex and longitudinal clinical features that are not obvious to clinicians and patients. These models predict the current stage and the future progression of diseases according to the latent patterns. AI-based prediction models provide new opportunities for clinicians to take proactive steps to prevent or mitigate potential health problems and identify patients who are at high risk of negative outcomes. For example, determining the course of chronic disease and displaying the information as a patient trajectory over time helps clinicians and patients to understand and manage progressive diseases. Powered by AI models, clinicians can use trajectory modeling of individual patients to better predict outcomes, enable earlier interventions, and modify risk factors to deliver targeted care. For example, the DisEase PrOgression Trajectory (DEPOT)[8], an evidence-driven, graph AI approach, is developed to model CKD progression trajectories and individualize clinical decision support (CDS). However, despite the promising performance, the black-box nature of AI models impedes clinical implementation as a CDS tool. It is challenging for clinicians to integrate AI results into a time-limited clinic appointment. In a recent study[9], among the 240 surveyed clinicians, only 43% were very familiar with machine learning. This study also shows incorrect machine learning-based treatment recommendations may adversely impact clinician treatment selections.

Visualization provides promising solutions to address the pressing needs[10] of incorporating AI models into clinical use. For example, modern computer visualization technologies provide intuitive illustrations of the predicted health trajectories and disease progression risks, as well as the linkage between EMR data and model predictions[11]. Meanwhile, explanatory visualization provides insights into which clinical indicators contribute to the prediction results for the specific patient. Furthermore, visual analytics (VA) has been proven to support chronic disease management with applied EMR data[12]. Therefore, visualization can be extremely useful for comprehending patient trajectories generated by AI or machine learning (ML) models. By presenting data in a graphical format, visualization helps to identify patterns and relationships that may not be immediately apparent from tabular data. Thus, visualization serves as a bridge to enhance the interpretation of complex AI/ML algorithms, making it easier for clinicians to understand how the algorithms derive a specific decision or prediction. For instance, the visualization of an AI algorithm could show how various features from a patient's EMR were weighted in predicting her health outcome.

To effectively visualize EMRs with their complex heterogeneity, Five Ws[13] presented a framework based on the who, what, when, where, and why of the data to tailor the visualization to the needs of healthcare users; CareVis[14] integrated patient data and clinical protocols to highlight deviations. Timeline, heatmap, and Sankey diagram-based interfaces are commonly used to visualize individual records, such as Lifelines[15], MediCoSpace[16], and Huang et al.[17]. However, analyzing cohorts or multiple patients requires space-efficient representation and aggerated algorithms (clustering, summaries, comparison[18]), for example, Gnaeus[19] combined AI and knowledge-assisted visualization to analyze cohort. In this work, we have utilized a timeline visual representation that allows for easy detection of abnormal indicator values and provides clues as to which indicators contribute to the trajectory[18, 19].

Some studies have explored the visualization of AI/ML models in clinical fields. RetainVis[20] was designed to understand Recurrent Neural Networks by providing an interactive interface to explore outputs and refine parameters; Vbridge[11]'s visualization incorporated feature attribution and model inversion to improve the trustworthiness of the AI model; DPVis[21] utilized hidden Markov models to pattern the patient disease progression pathways and presented them visually to identify milestones and intervention points; DG-Viz[22] combined deep learning and medical domain knowledge to enable exploration of risk prediction. Our work proposes intuitive visualizations for trajectories in both latent and realistic spaces, catering to both individual patients and groups.

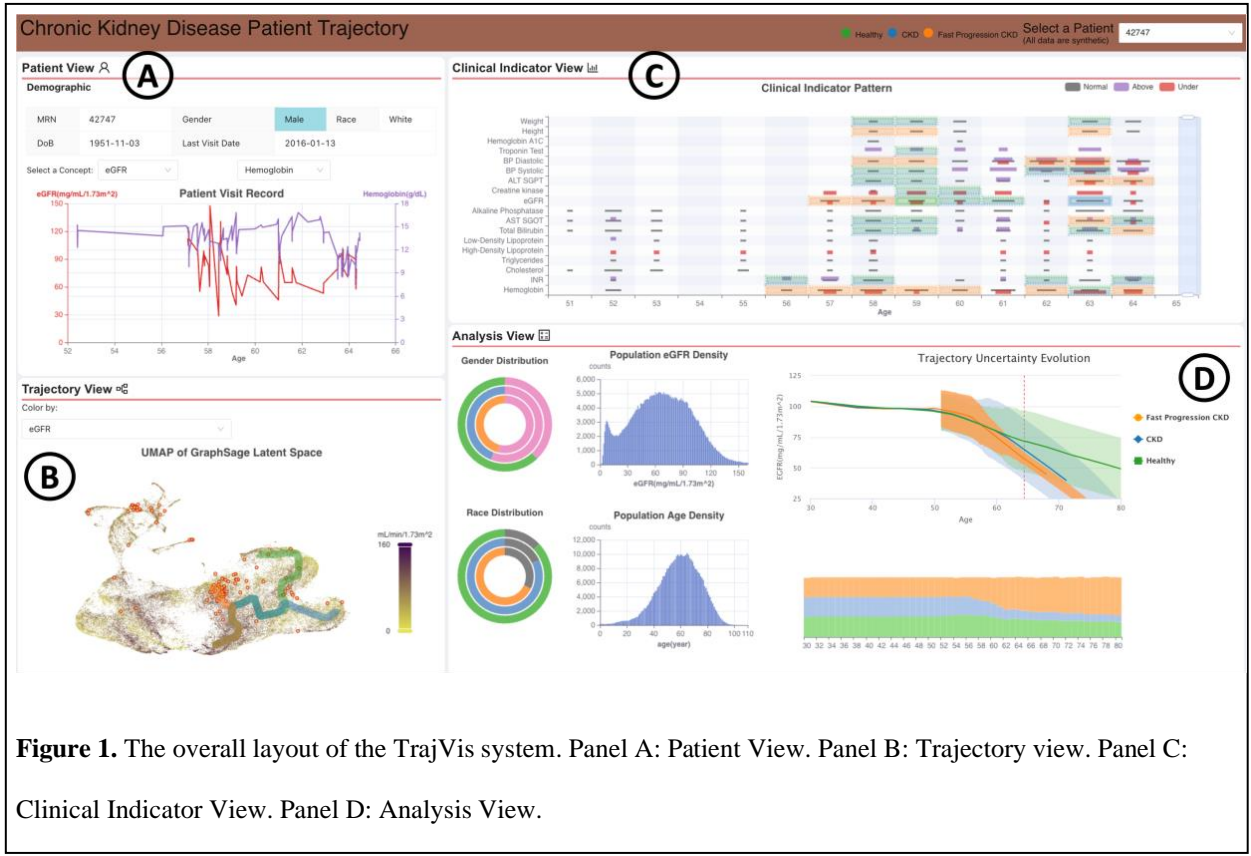

**Figure 1.** The overall layout of the TrajVis system. Panel A: Patient View. Panel B: Trajectory view. Panel C: Clinical Indicator View. Panel D: Analysis View.

In this work, we have developed a visual analysis system, TrajVis, to assist clinicians in comprehending AI models that offer health trajectory predictions, and to facilitate the application of AI models in daily clinical care. The DEPOT graph AI model[8] is used to demonstrate the functionalities of the TrajVis system. As shown in **Figure 1**, TrajVis displays the trends of essential clinical indicators for an individual patient (panel A), visualizes the learned latent representation of this patient's medical history within the training population and the learned trajectories (panel B), illustrates the longitudinal patterns of the patient's clinical visits, marks abnormal results of clinical indicators, highlights clinical evidence such as biomarkers and risk predictors linked to the AI-predicted disease progression fates (panel C), and demonstrates the predicted disease progression trajectory for individual patients (panel D). These functionalities of TrajVis support clinicians in effectively using AI/ML models to manage disease progression for individual patients. We further evaluate our system through a case study with clinicians and AI experts. The survey results show the effectiveness of our TrajVis system.

## 2. Methods

### 2.1 Requirement Analysis

The intended users include both clinicians and data scientists. The primary objective of TrajVis is to present a visual representation of the AI-predicted trajectory for individual patients, along with the clinical explanations of patient outcomes. We organized regular meetings involving two healthcare experts and two ML experts who possess expertise in machine learning, clinical research, and data visualization. These interactions categorize user requirements into the following tasks:

T1. Explore patient demographics and longitudinal visit records.

T2. Visualize latent representation of patient visits in context.

T3. Access the patient's medical history for visit pattern identification.

T4. Provide a visual representation of the predicted trajectory.

T5. Illustrate the contribution of visits and indicators to the predicted trajectory.

### 2.2 Data

TrajVis builds upon the deidentified EMR from the Atrium Health Wake Forest Baptist Translational Data Warehouse (WF-TDW) and the Indiana Network for Patient Care (INPC)[23] research database. The INPC data represents one of the largest health exchanges in the country and covers EMR data over more than 100 hospitals, 14,000 practices, and nearly 40,000 providers since the 1990s, which is managed by the Indiana Health Information Exchange (IHIE) [24] and Regenstrief Institute. Meanwhile, TrajVis embeds the DEPOT graph AI model as the computing component for revealing the CKD progression trajectories.

The EMR data have been harmonized according to our previously defined Common cLINic Index for Chronic Diseases (CLINIC) common data model (CDM)[25] to comprehensively characterize general health conditions: 1) reflecting health status and risks of chronic disease during aging; 2) preferring objective measurements over diagnosis; and 3) generally available in EMR. Essential clinical indices included in this work are listed in **Table 1**.

As described in our previous work[8], a total of 3,665,480 patients (79,434 from WF-TDW between 01/01/1993 and 06/30/2018, and 3,586,046 from INPC between 01/01/2016 and 05/31/2022), 34,320,143 clinical encounters (508,733 from WF-TDW and 33,811,410 from INPC), and 1,396,951,203 data elements (10,683,393 from WF-TDW and 1,386,267,810 from INPC) are included in this work. Among them, 2,050,537 (55.9%) are females, 484,172 (13.2%) are African Americans, and the median encounter ages are 60.1 (interquartile range or IQR: 49.5,70.0) and 60.59 (IQR: 45.31,72.01) for the WF-TDW and the INPC cohorts respectively.

**Table 1. CLINIC CDM and Essential Clinical Indices (ECI).** Italic: derived features.

**CLINIC CDM and Essential Clinical Indices**
- Demographics (3 features):
  - Sex
  - Race
  - *Age*
- Vitals (4 features):
  - Diastolic blood pressure
  - Systolic blood pressure
  - Height
  - Weight
- Laboratory tests (15 features):
  - Alanine aminotransferase (ALT/SGPT)
  - Aspartate aminotransferase (AST/SGOT)
  - Alkaline phosphatase (ALK)
  - Total cholesterol
  - Low density lipoprotein cholesterol (LDL)
  - High density lipoprotein cholesterol (HDL)
  - Creatine kinase
  - *Estimated glomerular filtrating rate (eGFR)*
  - Hemoglobin
  - Hemoglobin A1c (HbA1c)
  - Triglycerides
  - International normalized ratio of prothrombin time (INR)
  - Troponin

For demonstration purposes, de-identified synthetic longitudinal EMR records of 79,434 patients, 508,731 encounters, and 1,057,253 records were generated from the original data, with the dates randomly shifted for up to 6 months, and 10% encounter data replaced by similar encounters of different patients.

**2.3 Latent representation learning using DEPOT**

To identify possible disease progression trajectories, our first task is to build a latent representation from the input data. Briefly, an age-similarity graph is constructed with each node representing a clinical visit, each edge representing two visits that happened at similar ages (age difference less than 30 days), and the clinical features (from Essential Clinical Indices in **Table 1**) as the node attributes. The clinical features collected at a clinical visit $v_{s,t}$ for patient $s$ and visit $t$ are denoted as a numeric vector $\boldsymbol{c}_{s,t}$. After DEPOT projection, an 18-dimension latent representation $\boldsymbol{u}_{s,t}$ of the input data is learned and used for trajectory learning.

## 2.4 Trajectory learning using simple principal tree

Based on the latent representation, the disease progression trajectories are identified using the simple principal tree[27] for reversed graph embedding, with the latent features $\boldsymbol{u}_{s,t}$ are first projected to a 2-dimension space using the uniform manifold approximation and projection (UMAP)[26].

Specifically, the disease progression trajectories are represented as a principal minimum spanning tree $\mathcal{T} = \{\mathcal{V}, \mathcal{E}\}$, where $\mathcal{V} = \{v_1, \cdots, v_M\}$ are nodes known as "landmarks" that comprehensively represent gradually evolving clinical conditions, and $e_{i,j} \in \mathcal{E}$ represents the progression from one condition $v_i$ to its nearest condition $v_j$, where $v_i, v_j \in \mathcal{V}$. A specific landmark $v_i$ represents a set of similar clinical visits $\{u_{s,t}\}$. Therefore, each landmark $v_i$ can be annotated by corresponding clinical features according to the clinical visits associated with it. For example, the age that a landmark $v_i$ represents can be defined as the median age of the associated clinical visits $\{v_{s,t}\}|_{v_i}$. We further define the fork nodes as nodes of degree 3. The tree can therefore be segmented into branches, with each branch defined as the smallest chain between two forks or between a fork and a terminal. The longitudinal direction of each branch is determined by the overall aging trends of its landmark nodes. For example, for a branch $\mathcal{T}_B = \{\mathcal{V}_B, \mathcal{E}_B\}$ with nodes $\{v_1, \cdots, v_B\}$, the direction of this branch is defined as $v_1 \to v_B$ if $\sum_{i=1}^{B}(age_{i+1} - age_i) > 0$. Totally 100 landmarks are used in trajectory learning.

The identified trajectories are then evaluated for relevance with CKD progression based on the Pearson correlation between the order of the nodes on each branch and their corresponding eGFR values. For visualization, the learned trajectories are smoothed through robust locally weighted regressions and smoothing scatterplots (robust LOWESS)[28] with a span fraction of 2/3 over three regression iterations. The disease progression trajectory of an individual patient at a specific age is then determined according to the attributions of the patient's clinical visits along the learned trajectories. For a specific patient $s$ at age $a_t$, the possibility of this patient belonging to a discovered trajectory $\mathcal{T}_b$, i.e., $P_{s,b,a_t}$, is determined as: $P_{s,b,a_t} = \frac{|v_{s,\mathcal{T}_b,a}|a \leq a_t|}{|v_{s,\cdot,a}|a \leq a_t|}$, where $v_{p,b,a}$ represents the clinical visit of patient $s$ at age $a$ attributed to the trajectory $\mathcal{T}_b$.

## 2.6 Explainable visualization of the DEPOT model

Clinical features associated with the trajectories are analyzed through enrichment analysis at the cohort level and visualized at the individual patient level to elucidate which clinical features contribute to AI/ML predictions. Specifically, clinical features of the visits before a fork node and associated with a specific trajectory after the fork node are defined as the predictors of the trajectory; clinical features concurrently associated with a trajectory are defined as its markers. To examine whether a numeric feature $c$ is the predictor of a terminal trajectory $\mathcal{T}_B$, Welch's t-test $t(c_{s,t}|t < v_{f,\mathcal{T}_B}, s \sim \mathcal{T}_B, c_{s,t}|t < v_{f,\mathcal{T}_B}, s \sim \mathcal{T}_B^C)$ is used, where $c_{s,t}$ is the numeric result of clinical feature $c$ of patient $s$ at visit $t$, $v_{f,\mathcal{T}_B}$ is the fork landmark that separates trajectory $\mathcal{T}_B$ from the rest graph, $t < v_{f,\mathcal{T}_B}$ represents visits that happened before the trajectory $\mathcal{T}_B$, $\mathcal{T}_B^C$ represents other trajectories after the fork landmark $v_{f,\mathcal{T}_B}$, and $s \sim \mathcal{T}_B$ means the visits of patient $s$ that are attributed to branch $\mathcal{T}_B$. Chi-square tests are used for categorical clinical features. The rationale is that, if a clinical feature before the branching point demonstrates significant differences between patients who progress toward the trajectory $\mathcal{T}_B$ and those who do not, this feature provides clinical insights related to future chances of trajectory $\mathcal{T}_B$. Clinical markers can be identified similarly, with $t(c_{s,t}|t > v_{f,\mathcal{T}_B}, s \sim \mathcal{T}_B, c_{s,t}|t > v_{f,\mathcal{T}_B}, s \sim \mathcal{T}_B^C)$ for numeric features, and chi-square tests for categorical features. The Benjamini-Hochberg false discovery rates (FDR)[29] is used to adjust for multiple comparisons, with FDR <0.05 considered significant.

## 3. Visual Analytics System: TrajVis

We implement the system's backend with Python and Flask and the frontend with React and Echart. The architecture and workflow are illustrated in **Figure 2**. The TrajVis system workflow, which runs at the backend, is composed of four components: 1) The data component which is composed of the harmonized WF-TDW and INPC EMR data. 2) The AI/ML model component. The DEPOT model is available as a containerized image or as a package. Other trajectory learning models, such as the group-based trajectory models (GBTM) and dynamic time warping, can be incorporated into this component. 3) Patient trajectory. Patients' clinical visits can be projected and visualized in the latent space in the context of the three established CKD trajectories (healthy, late progression, and early progression trajectories). The latent space is visualized using UMAP. 4) The interpretation component provides the predictors and markers associating the EMR data of a patient to the three trajectories, as well as the possibility that the patient is associated with each trajectory. The React-based frontend interface of TrajVis communicates with the backend components via a RESTful API interface. A demonstration version of TrajVis is available at https://trajvis.sulab.io.

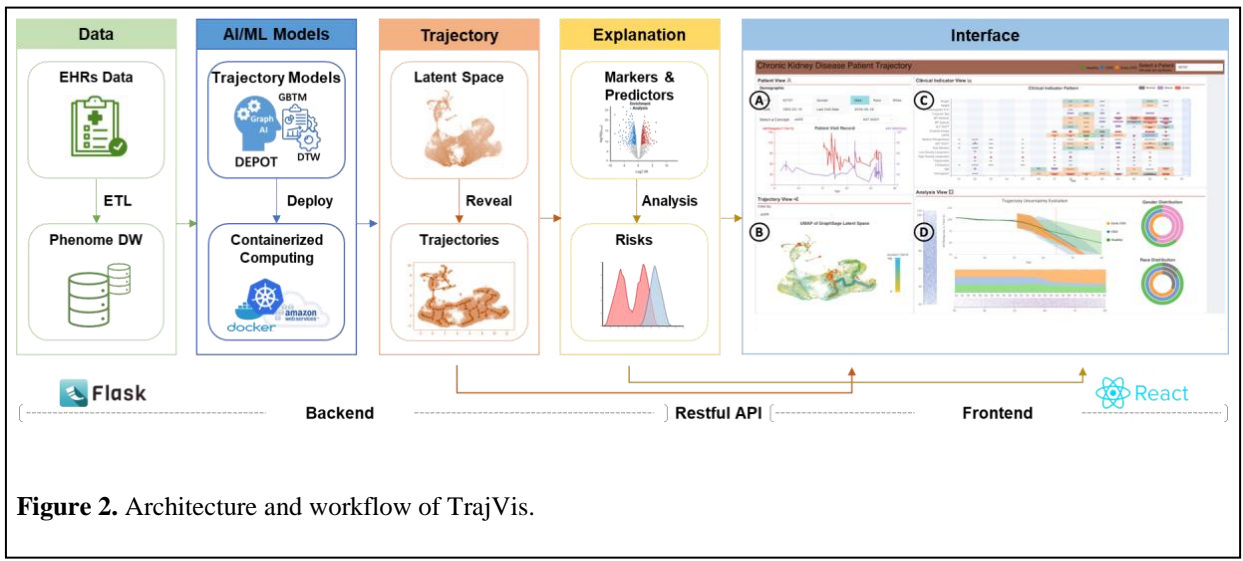

**Figure 2.** Architecture and workflow of TrajVis.

The TrajVis Interface comprises the following view panels (**Figure 1**):

### 3.1 Patient Profile View

The patient profile view (**Figure 1A**) provides basic demographic and pathologic information of patients.

Clinicians can select a patient from the dropdown menu, examine different clinical indicators such as the estimated glomerular filtration rate (eGFR) or serum hemoglobin levels from the dropdown menu at the left, and compare up to two indicator trends in one chart. This feature enables clinicians to visualize the relationships between two clinical indicators with respect to the same time axis, such as cholesterol and blood pressure.

### 3.2 Trajectory View

The UMAP visualization illustrates the DEPOT latent representation of clinical visit data for the entire cohort. **Figure 1b** displays the three learned CKD progression trajectories, with the healthy, CKD progression, and fast progression marked in green, blue, and orange trend lines. Users can select different clinical features (eGFR, age, etc.) or color themes from the dropdown menu to annotate the latent representations. Clinical meanings of clinical features are color-coded. For example, deeper colors represent advancing age along the trajectory. The medical history of the selected patient during each visit is highlighted as scattered circles, allowing clinicians to view this patient in the context of the population.

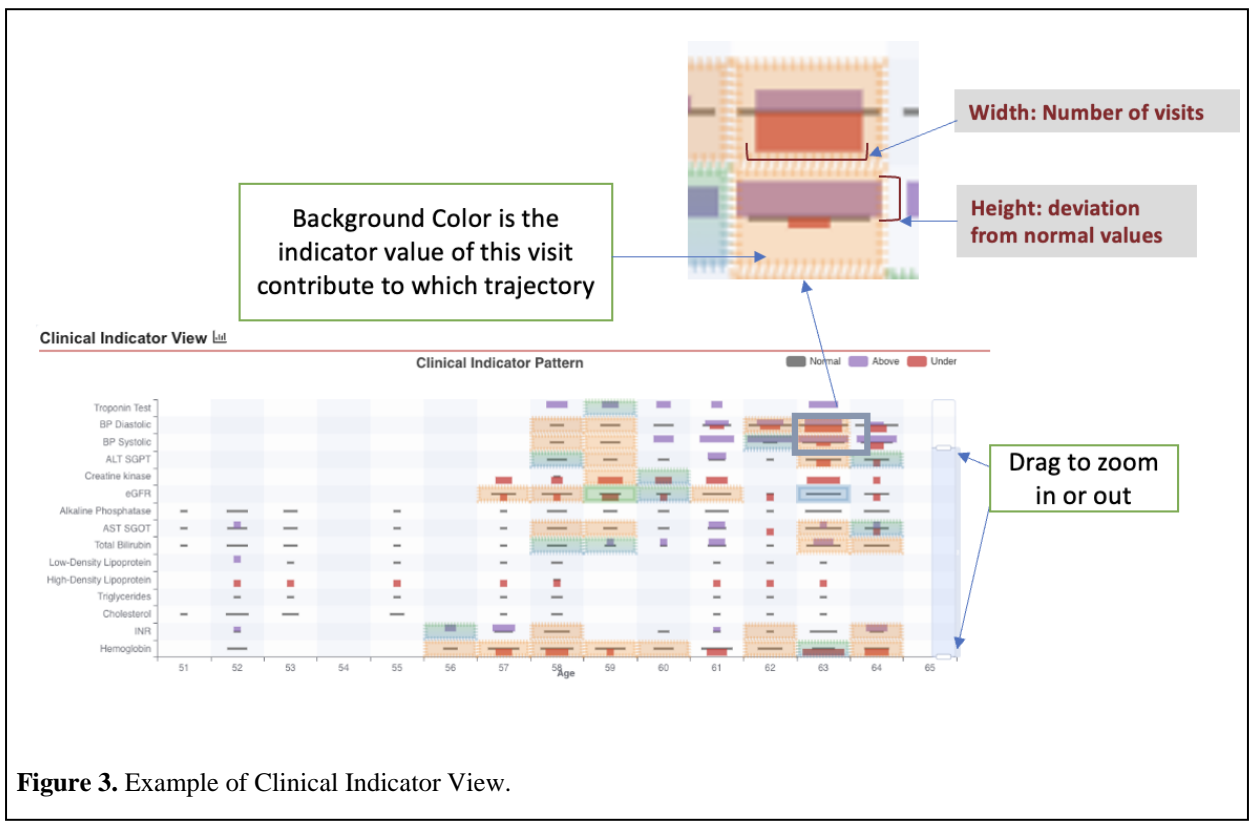

**Figure 3.** Example of Clinical Indicator View.

This view provides an intuitive visualization of the topological structure of clinical data, summarizing patient visits, and aids clinicians in comprehending complex information in a user-friendly manner. The two-dimensional format of the latent space representation, derived from the high-dimensional indicator values, assists clinicians in understanding the origins of trajectories and the correlations between trajectories and clinical features. For data scientists, this view facilitates confirming the presence of patterns, identifying those patterns associated with CKD, and evaluating the effectiveness of the trajectory in capturing these patterns.

### 3.3 Clinical Indicator View

This view provides clinicians with an overview of a patient's visit patterns, trends in vital signs and laboratory test results, and clinical evidence used by DEPOT for predicting this patient's CKD progression trajectory. Markers and predictors associated with the three trajectories are labeled in the corresponding colors. We use symbols to mark the presence of visits across the ages (as shown in **Figure 3**). Each symbol has three vertical segments: normal values, above-normal values, and under-normal values. The width of these segments reflects the number of values, while the height indicates the deviation from normal values. Additionally, hierarchical clustering of data availability is applied to organize the clinical indicators, assisting clinicians in holistically comprehending longitudinal patterns across different lab tests and identifying clinical concerns behind such patterns. This visualization design is based on clinicians' inputs on their need for an overview of data availability, the timing of emerging concerns, which of these concerns are linked to CKD, and which laboratory test values correlate with the CKD trajectory. By staying informed about health trends, clinicians can personalize healthcare to their patients.

### 3.4 The Analysis View

The visualization includes a line chart depicting eGFR levels across different age groups for each trajectory cohort. Around the line in the line chart, a color-coded band conveys the uncertainty associated with transitioning to each trajectory over time. Below the line chart, a possibility band illustrates the likelihood of a given patient adhering to each trajectory throughout the timeline. Additionally, two pie charts showcase the gender and race distribution among the three trajectories, while two density plots outline the age and eGFR distribution in the population. This comprehensive view empowers clinicians to scrutinize eGFR patterns within each trajectory cohort and track the evolution of the probability of transitioning to each trajectory.

## 4. Evaluation

We use a case study to demonstrate how TrajVis can facilitate clinicians to intuitively understand the results of the DEPOT graph AI model and use such AI/ML models to manage patients' CKD progression. A questionnaire-based user survey is performed to evaluate the design and functionalities of TrajVis.

### 4.1 Case Study

A clinician assesses whether TrajVis can help clinicians identify different CKD progression trajectories, investigate patients' medical history, and understand which clinical features were associated with the different fates.

The clinician examined a white male patient (identifier: 42747) and a female African American (identifier: 58314). As demonstrated in the Patient Profile View (**Figure 4A**), at age 57.6, patient 42747 showed a temporary eGFR decline below 60mg/mL/1.73m$^2$, suffered an AKI episode at age 58.5, and quickly recovered within three months. However, this patient later suffered another AKI episode at age 59.3, leading to sustained kidney injury (CKD stage G3) thereafter. In contrast, Patient 58314 experienced an AKI episode at age 48.2, exhibited resilient kidney function, and recovered from five acute kidney injury episodes in the following decade. The clinician concludes that the Profile View serves as a convenient panel for assessing the longitudinal trends of essential clinical indicators. Furthermore, the clinician agrees that eGFR alone is not sufficient for monitoring patients' risk of kidney function decline, as evidenced by the divergent trajectory of patient 42747, who appeared to recover after the initial AKI event but subsequently demonstrated rapid progression of kidney disease, and patient 58314, who suffered frequent AKI incidents

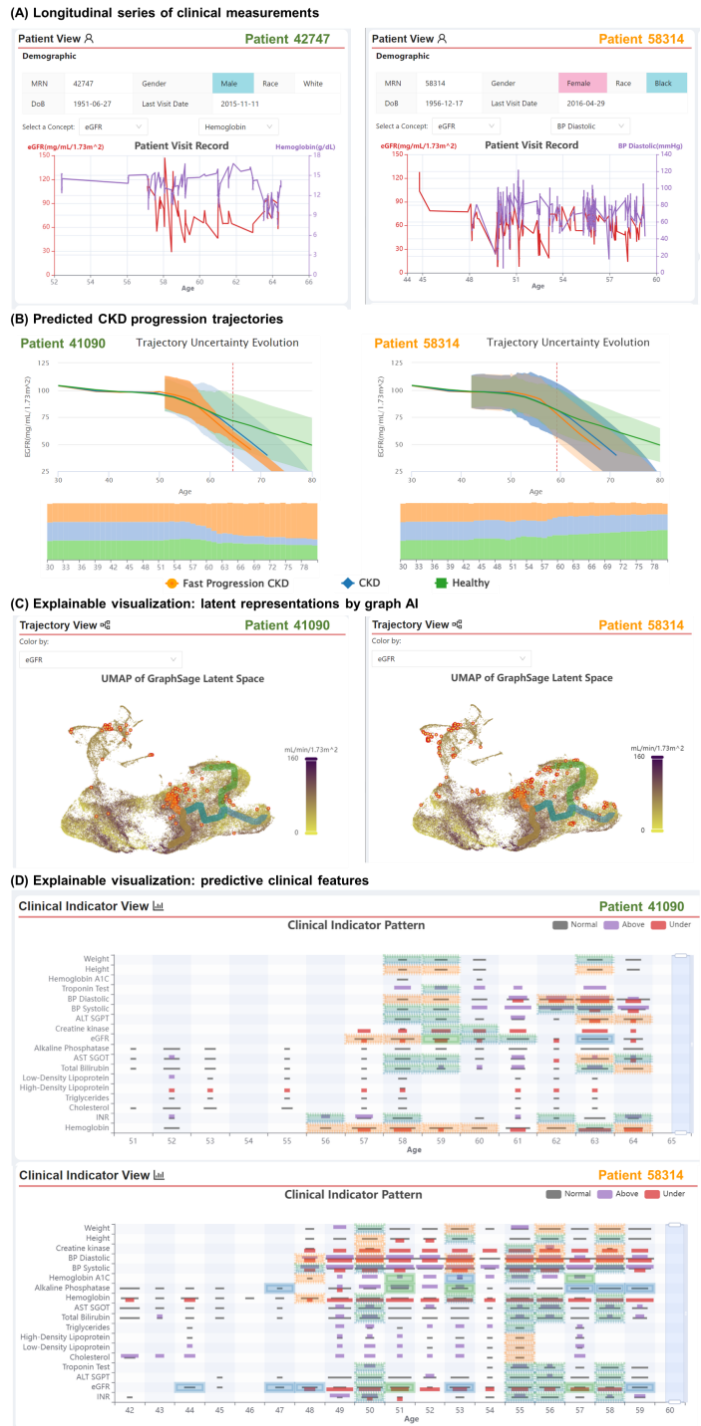

**Figure 4.** Case study of two typical patients. (pseudo MRN numbers 42747 and 58314). (A) Longitudinal series of clinical measurements. (B) Predicted CKD progression trajectories. (C) Explainable visualization of the latent representations by graph AI. (D) Explainable visualization of the predictive clinical features. Two patients demonstrated in this case study were with the pseudo-MRN numbers 42747 and 58314.

for over a decade yet remained resilience.

The Analysis View (**Figure 4B**) illustrates the prediction results from the graph AI model, DEPOT, for both patents. The clinician observes that as early as age 47, the DEPOT model predicted an escalating risk of patient 42747 following the Fast Progression trajectory. This early prediction could have provided nephrologists and the patient with two years for timely interventions. In contrast, the clinician notes that despite numerous acute kidney injury incidents, Patient 58314 exhibited stable kidney function. Those observations demonstrate the intuitive and informative characteristics of Analysis View. Subsequently, the clinician consults the Trajectory View (**Figure 4C**). The color-coded annotations of eGFR and age facilitate the holistic visualization of clinical information and the understanding of the identified trajectories, with the clinical visits of patient 42747 predominately aligning with the Fast Profession trajectory, while the clinical visits of patient 58314 undetermined. This explained the predictions of the disease progression trajectories of the two patients. Finally, the clinician examines the Clinical Indicator View for clinical indicators and their roles in predicting patients' disease progressions. The low hemoglobin level of patient 42747 starting at age 57 is indicated as a potential contributor to or sequelae from the patient's rapid progression of kidney disease. The primary risk predictor for patient 58314 was unstable diastolic blood pressure, which was further confirmed by checking the trends of this factor in the Patient View (**Figure 4A**, right panel). Additional minor contributors included her comorbidities of obesity and well-controlled diabetes mellitus. The

> **Box 1.** Questionnaire of user experience.
>
> *Questions:*
>
> - *How intuitive are the layouts and functionalities of the system?*
> - *How helpful do you find the multiple interactive views in aiding you to identify longitudinal patterns of clinical indicators?*
> - *How easy is it for a new user to learn and use the tool?*
> - *Compared to the current tools you are using, does this tool help you understand the trajectory learned from an artificial intelligence model?*
> - *Could you please discuss the potential uses and benefits of this tool?*
> - *Overall satisfaction.*

clinician concludes that the Clinical Indicator View of the major clinical measures that are associated with the AI's predictions, accompanied by the Patient View of the clinical measures, allows clinicians to understand the clinical evidence supporting the AI model and provides insights into potential interventions. In summary, the TrajVis tool offers a comprehensive and informative interface for examining the longitudinal trends of an individual patient's clinical indicators, explaining how the DEPOT graph AI model predicts trajectories, and shedding light on potential personalized interventions.

## 4.2 Questionnaire-based User Survey

We conduct a user experience survey consisting of six questions (**Box 1**) to test the hypothesis that TrajVis is satisfactory concerning design and functionality. A total of 17 participants are recruited to the survey from Nephrology at IU Health and Purdue Polytechnic Institute through convenience sampling. The demographic information of the participants is shown in **Table 2**. The participants are first provided a brief tutorial and allowed 15 minutes to explore the system. Subsequently, they are asked to examine the design rationale, the functionality of the tool, and the difficulty in learning. The participants rate their satisfaction levels on a scale from 1 (highly unsatisfied) to 7 (highly satisfied) and provide open-ended feedback.

**Table 2.** Characteristics of surveyed users.

| Characteristics | N | % |
| --- | --- | --- |
| Overall | 17 | |
| **Gender** | | |
| Male | 10 | 59% |
| Female | 7 | 41% |
| **Age** | | |
| 18-25 | 2 | 12% |
| 26-35 | 9 | 53% |
| 36-45 | 3 | 17% |
| 46-55 | 2 | 12% |
| Over 55 | 1 | 6% |
| **Healthcare Experience** | | |
| No experience | 3 | 17% |
| < 5 years | 8 | 47% |
| 5-15 years | 3 | 17% |
| 16-25 years | 2 | 12% |
| > 25 years | 1 | 6% |
| **Computing Experience** | | |
| Yes | 11 | 65% |
| No | 6 | 35% |

The survey results (**Figure 5**) show that 12 participants are very satisfied (score >= 6) with the integration of AI by TrajVis. Individuals with computational expertise rate TrajVis as intuitive, helpful, and easy. Even for clinicians without a computing background, half are comfortable with using TrajVis to integrate AI into clinical care. Some of the participants' opinions are shown in **Box 2**.

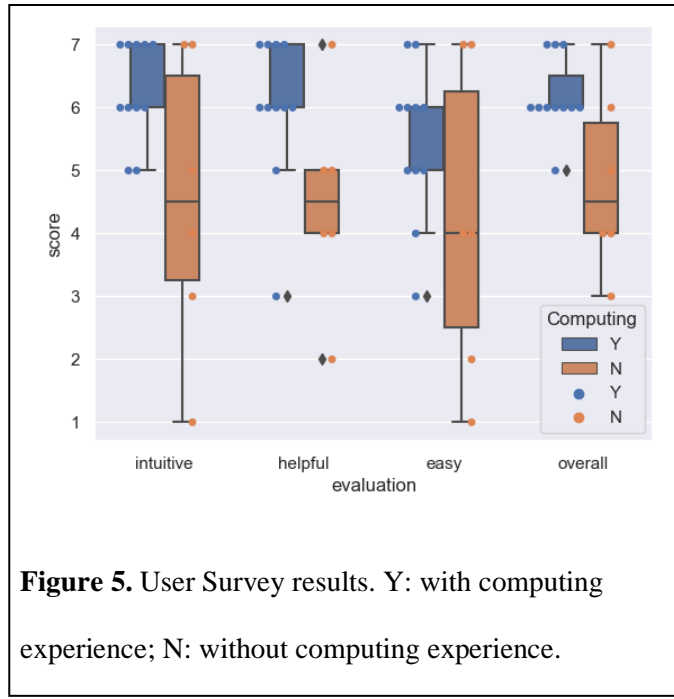

**Figure 5.** User Survey results. Y: with computing experience; N: without computing experience.

## 5. Conclusion

In this work, we develop TrajVis to address the key challenge in clinical visualization and in

**Box 2.** Participants' Feedback.

A clinician with over 25 years of experience: *It appears that the tool will be best used to identify the patients with the highest disease progression risk that escapes common clinical surveillance.*

A clinician with over 15 years of experience: *In the future this tool could be useful to screen populations of patients for the need for referral into renal clinic.*

A clinician with 5-15 years of experience: *Plotting trajectories of kidney function decline is something nephrologists do every day in practice, what is great about TrajVis is that the program predicts future possibilities for our patients based on all available information.*

A data scientist with 10 years of experience: *This tool allows complex data easier to understand. By presenting data in a graphical format, it is easier for users to identify the trajectory patterns and trends that are difficult to discern from raw data. Another potential benefit is that it can be used for exploratory analysis, enabling users to interact with data in real-time and explore different scenarios. Potential usage: (1) show the trajectory of the EMR data. (2) show genomics data like 10x benefits: visualize and aid the pathologists to better understand the datasets.* [Note: 10x means the genomics data generated from 10x Genomics' platform, including single-cell RNA-Seq, single-cell ATAC-Seq, Visium Spatial, and Xenium in Situ. Such technologies have recently been used at IU Health for pathologic nephrology. Pathologists need a dashboard to provide a comprehensive medical background when examining 10x genomics data.]

A data scientist with over 15 years of experience: *It presents the result of AI model, helping people understand the usage of AI.*

the use of AI/ML models to manage the progression of chronic diseases, with the DEPOT graph AI model and personalized CKD management as the use case. The case study demonstrates the feasibility and effectiveness of the system. TrajVis bridges the gap between the fast-growing AI/ML modeling and the clinical use of such models for personalized and precision management of chronic diseases.

**Acknowledgment**

We acknowledge Regenstrief Institute Inc.'s sourcing of data for this project.

**Funding**

J.S., M.E., P.Z., and X.L. are financially supported by the National Library of Medicine of the National Institute of Health (R01LM013771). J.S. is also supported by the Indiana University Precision Health Initiative and the Indiana University Melvin and Bren Simon Comprehensive Cancer Center Support Grant from the National Cancer Institute (P30CA 082709). Q.S. is supported by the National Institute of General Medical Sciences of the National Institutes of Health (R35GM151089).

**Data availability**

The synthetic demo data is available on GitHub (https://github.com/Su-informatics-lab/trajvis). The electronic medical records at WFBH and INPC are controlled data and, thus, are not available.

**Code availability**

The TrajVis data analysis and modeling are provided as an open-source package on GitHub (https://github.com/Su-informatics-lab/trajvis).

The DEPOT method is provided as an open-source package on GitHub (https://github.com/jing-su/depot).

**The role of the Institutional Review Boards**

The study has been covered under approved protocols by the Wake Forest School of Medicine Institutional Review Board and the Indiana University Institutional Review Board.

# References

1. Levin, A. et al. Kidney Disease: Improving Global Outcomes (KDIGO) CKD Work Group. KDIGO 2012 clinical practice guideline for the evaluation and management of chronic kidney disease. *Kidney International Supplements* **3**, 1-150 (2013).

2. Vassalotti, J.A. et al. Practical Approach to Detection and Management of Chronic Kidney Disease for the Primary Care Clinician. *Am J Med* **129**, 153-162 e157 (2016).

3. Gilmore, J. KDOQI clinical practice guidelines and clinical practice recommendations--2006 updates. *Nephrol Nurs J* **33**, 487-488 (2006).

4. Levey, A.S. et al. The definition, classification, and prognosis of chronic kidney disease: a KDIGO Controversies Conference report. *Kidney Int* **80**, 17-28 (2011).

5. Soranno, D.E. et al. Artificial Intelligence for AKI!Now: Let's Not Await Plato's Utopian Republic. *Kidney360* **3**, 376-381 (2022).

6. Bajaj, T. & Koyner, J.L. Artificial Intelligence in Acute Kidney Injury Prediction. *Adv Chronic Kidney Dis* **29**, 450-460 (2022).

7. Sanmarchi, F. et al. Predict, diagnose, and treat chronic kidney disease with machine learning: a systematic literature review. *J Nephrol* **36**, 1101-1117 (2023).

8. Song, Q. et al. DEPOT: graph learning delineates the roles of cancers in the progression trajectories of chronic kidney disease using electronic medical records. *medRxiv*, 2023.2008.2013.23293968 (2023).

9. Jacobs, M. et al. How machine-learning recommendations influence clinician treatment selections: the example of the antidepressant selection. *Transl Psychiatry* **11**, 108 (2021).

10. Choudhury, A., Renjilian, E. & Asan, O. Use of machine learning in geriatric clinical care for chronic diseases: a systematic literature review. *JAMIA Open* **3**, 459-471 (2020).

11. Cheng, F. et al. Vbridge: Connecting the dots between features and data to explain healthcare models. *IEEE Transactions on Visualization and Computer Graphics* **28**, 378-388 (2021).

12. Feller, D.J. et al. A visual analytics approach for pattern-recognition in patient-generated data. *J Am Med Inform Assoc* **25**, 1366-1374 (2018).

13. Zhang, Z. et al. The five Ws for information visualization with application to healthcare informatics. *IEEE Trans Vis Comput Graph* **19**, 1895-1910 (2013).

14. Aigner, W. & Miksch, S. CareVis: integrated visualization of computerized protocols and temporal patient data. *Artif Intell Med* **37**, 203-218 (2006).

15. Plaisant, C. et al. LifeLines: using visualization to enhance navigation and analysis of patient records. *Proceedings / AMIA ... Annual Symposium. AMIA Symposium*, 76-80 (1998).

16. van der Linden, S., Sevastjanova, R., Funk, M. & El-Assady, M. MediCoSpace: Visual Decision-Support for Doctor-Patient Consultations using Medical Concept Spaces from EHRs. *ACM Transactions on Management Information Systems* **14**, 1-20 (2023).

17. Huang, C.W. et al. A richly interactive exploratory data analysis and visualization tool using electronic medical records. *BMC Med Inform Decis Mak* **15**, 92 (2015).

18. Guo, Y. et al. Survey on visual analysis of event sequence data. *IEEE Transactions on Visualization and Computer Graphics* **28**, 5091-5112 (2021).

19. Federico, P. et al.  79-83 (

20. Kwon, B.C. et al. Retainvis: Visual analytics with interpretable and interactive recurrent neural networks on electronic medical records. *IEEE transactions on visualization and computer graphics* **25**, 299-309 (2018).

21. Kwon, B.C. et al. DPVis: Visual analytics with hidden markov models for disease progression pathways. *IEEE transactions on visualization and computer graphics* **27**, 3685-3700 (2020).

22. Li, R., Yin, C., Yang, S., Qian, B. & Zhang, P. Marrying Medical Domain Knowledge With Deep Learning on Electronic Health Records: A Deep Visual Analytics Approach. *J Med Internet Res* **22**, e20645 (2020).

23. McDonald, C.J. et al. The Indiana network for patient care: a working local health information infrastructure. An example of a working infrastructure collaboration that links data from five health systems and hundreds of millions of entries. *Health Aff (Millwood)* **24**, 1214-1220 (2005).

24. Overhage, J.M. & Kansky, J.P. in Health Information Exchange (Second Edition). (ed. B.E. Dixon) 471-487 (Academic Press, 2023).

25. Wang, L. et al. Progression of chronic kidney disease in African American with type 2 diabetes mellitus using topology learning in electronic medical records. *bioRxiv* (2018).

26. McInnes, L., Healy, J. & Melville, J. arXiv:1802.03426 (2018).

27. Mao, Q., Yang, L., Wang, L., Goodison, S. & Sun, Y. in Proceedings of the 2015 SIAM International Conference on Data Mining (SDM) 792-800 (Society for Industrial and Applied Mathematics, 2015).

28. Cleveland, W.S. Robust locally weighted regression and smoothing scatterplots. *J Am Stat Assoc* **74**, 829-836 (1979).

29. Benjamini, Y. & Hochberg, Y. Controlling the False Discovery Rate - a Practical and Powerful Approach to Multiple Testing. *J Roy Stat Soc B Met* **57**, 289-300 (1995).